\documentclass[a4paper,12pt]{article}
\usepackage{amsmath}
\usepackage{amssymb}
\usepackage{amsfonts}
\usepackage{amssymb,amsmath}
\usepackage{cancel}
\usepackage{graphicx}
\usepackage{epsfig}
\usepackage{verbatim}
\usepackage{fleqn}
\usepackage[footnotesize]{caption}
\usepackage{mathrsfs}

\def\beq{\begin{equation}}
\def\eeq{\end{equation}}
\def\bea{\begin{eqnarray}}
\def\eea{\end{eqnarray}}

\def\<{\left\langle}
\def\>{\right\rangle}

\newcommand{\bc}{\begin{center}}
\newcommand{\ec}{\end{center}}
\newcommand{\bd}{\begin{displaymath}}
\newcommand{\ed}{\end{displaymath}}
\newcommand{\be}{\begin{equation}}
\newcommand{\ee}{\end{equation}}
\newcommand{\ba}{\begin{array}}
\newcommand{\ea}{\end{array}}
\newcommand{\bt}{\begin{tabular}}
\newcommand{\et}{\end{tabular}}

\newcommand{\ds}{\displaystyle}

\begin{document}

\bibliographystyle{OurBibTeX}

\begin{titlepage}

\begin{flushright}
SHEP-08-37\\
\end{flushright}

\begin{center}
{ \sffamily \Large Predictions of the Constrained Exceptional
Supersymmetric Standard Model }
\\[8mm]
P.~Athron$^{a}$,
S.F.~King$^{b}$,
D.J.~Miller$^{c}$,
S.~Moretti$^{b}$
and
R.~Nevzorov$^{c}$\footnote{On leave of absence from the Theory Department,
ITEP, Moscow, Russia
}\\[3mm]
{\small\it
$^a$ Institut f\"ur Kern- und Teilchenphysik, TU Dresden, D-01062, Germany,\\[2mm]
$^b$ School of Physics and Astronomy, University of Southampton,\\
Southampton, SO17 1BJ, U.K.\\[2mm]
$^c$ Department of Physics and Astronomy, University of Glasgow,\\
Glasgow, G12 8QQ, U.K.}
\\[1mm]
\end{center}
\vspace*{0.75cm}

\begin{abstract}
\noindent
We discuss the predictions of a constrained version of the exceptional
supersymmetric standard model (cE$_6$SSM), based on a universal high
energy soft scalar mass $m_0$, soft trilinear coupling $A_0$ and soft
gaugino mass $M_{1/2}$. We predict a supersymmetry (SUSY) spectrum
containing a light gluino, a light wino-like neutralino and chargino
pair and a light bino-like neutralino, with other sparticle masses
except the lighter stop being much heavier.  In addition, the
cE$_6$SSM allows the possibility of light exotic colour triplet charge
$1/3$ fermions and scalars, leading to early exotic physics signals at
the LHC.  We focus on the possibility of a $Z'$ gauge boson with mass
close to 1 TeV, and low values of $(m_0,M_{1/2})$, which would
correspond to an LHC discovery using ``first data'', and propose a set
of benchmark points to illustrate this.
\end{abstract}

\end{titlepage}
\newpage
\setcounter{footnote}{0}

\section{Introduction}
The minimal supersymmetric standard model (MSSM) \cite{Chung:2003fi}
provides a very attractive supersymmetric extension of the standard
model (SM). Its superpotential contains the bilinear term $\mu
{H}_d{H}_u$, where ${H}_{d,u}$ are the two Higgs doublets which
develop vacuum expectation values (VEVs) at the weak scale and $\mu$
is the supersymmetric Higgs mass parameter which can be present before
SUSY is broken.  However, despite its attractiveness, the MSSM suffers
from the $\mu$ problem: one would naturally expect $\mu$ to be either
zero or of the order of the Planck scale, while, in order to get the
correct pattern of electroweak symmetry breaking (EWSB), $\mu$ is
required to be in the TeV range.

It is well known that the $\mu$ term of the MSSM can be generated
effectively by the low energy VEV of a singlet field $S$ via the
interaction $\lambda SH_d H_u$.  However, although an extra singlet
superfield seems like a minor modification to the MSSM, which does no
harm to either gauge coupling unification or neutralino dark matter,
its introduction leads to an additional accidental global $U(1)$
(Peccei-Quinn (PQ) \cite{Peccei:1977hh}) symmetry which will result in
a weak scale massless axion when it is spontaneously broken by
$\langle S\rangle$ \cite{Fayet:1974fj}. Since such an axion has not
been observed experimentally, it must be removed somehow. This can be
done in several ways resulting in different non-minimal SUSY models,
each involving additional fields and/or parameters
\cite{nmssm,othernmssm}. For example, the classic solution to this
problem is to introduce a singlet term $S^3$, as in the
next-to-minimal supersymmetric standard model (NMSSM)~\cite{nmssm},
which reduces the PQ symmetry to the discrete symmetry $Z_3$. However
the subsequent breaking of a discrete symmetry at the weak scale can
lead to cosmological domain walls which would overclose the Universe.

A cosmologically safe solution to the axion problem of singlet models,
which we follow in this Letter, is to promote the PQ symmetry to an
Abelian $U(1)'$ gauge symmetry \cite{Fayet:1977yc}.  The idea is that
the extra gauge boson will eat the troublesome axion via the Higgs
mechanism resulting in a massive $Z'$ at the TeV scale. The necessary
$U(1)'$ gauge group could be a relic of the breaking of some unified
gauge group at high energies. Recall that the unification of gauge
couplings in SUSY models allows one to embed the gauge group of the SM
into Grand Unified Theories (GUTs) based on simple gauge groups such
as $SU(5)$, $SO(10)$ or $E_6$. In particular the $E_6$ symmetry can be
broken to the rank--5 subgroup $SU(3)_C\times SU(2)_L\times
U(1)_Y\times U(1)'$ where in general $U(1)'=U(1)_{\chi}
\cos\theta+U(1)_{\psi} \sin\theta$ \cite{E6}, and the two anomaly-free
$U(1)_{\psi}$ and $U(1)_{\chi}$ symmetries originate from the
breakings $E_6\to$ $SO(10)\times U(1)_{\psi}$, $SO(10)\to$
$SU(5)\times$ $U(1)_{\chi}$ (for recent review see
\cite{U(1)-review}).

Within the class of $E_6$ models there is a unique choice of Abelian
gauge group that allows zero charges for right-handed neutrinos and
thus large Majorana masses and a high scale see-saw mechanism. This is
the $U(1)_{N}$ gauge symmetry given by $\theta=\arctan\sqrt{15}$, and
defines the so-called exceptional supersymmetric standard model
(E$_6$SSM) \cite{King:2005jy}. The extra $U(1)_{N}$ gauge symmetry
survives to low energies and forbids a bilinear term $\mu {H}_d {H}_u$
in the superpotential but allows the interaction $\lambda S H_d
H_u$. At the electroweak (EW) scale. the scalar component of the SM
singlet superfield $S$ acquires a non-zero VEV, $\langle S
\rangle=s/\sqrt{2}$, breaking $U(1)_N$ and yielding an effective
$\mu=\lambda s/\sqrt{2}$ term.  Thus the $\mu$ problem in the E$_6$SSM
is solved in a similar way to the NMSSM, but without the accompanying
problems of singlet tadpoles or domain walls. In this model low energy
anomalies are cancelled by complete 27 representations of $E_6$
which survive to low energies, with $E_6$ broken at the high energy
GUT scale.

In this Letter we discuss some of the predictions of particular
relevance to the LHC from a constrained version of the E$_6$SSM
(cE$_6$SSM), based on a universal high energy soft scalar mass $m_0$,
soft trilinear coupling $A_0$ and soft gaugino mass $M_{1/2}$.  Our
primary focus is on the most urgent regions of parameter space which
involve low values of $(m_0,M_{1/2})$ and low $Z'$ gauge boson masses
which would correspond to an early LHC discovery using ``first data''.
To illustrate these features we propose and discuss a set of ``early
discovery'' benchmark points, each associated with a $Z'$ gauge boson
mass around 1 TeV and $(m_0,M_{1/2})$ below 1 TeV, which would lead to
an early indication of the cE$_6$SSM at the LHC.  We find a SUSY
spectrum consisting of a light gluino of mass $\sim M_3$, a light
wino-like neutralino and chargino pair of mass $\sim M_2$, and a light
bino-like neutralino of mass $\sim M_1$, where $M_i$ are the low
energy gaugino masses, which are typically driven small by
renormalisation group (RG) running.  Sfermions are generally heavier,
but there can be an observable top squark. There may also be light
exotic colour triplet charge $1/3$ fermions and scalars, whose masses
are controlled by independent Yukawa couplings.  Some first results
have already been trailed at conferences \cite{Athron:2008np} and a
longer paper containing full details of the analysis is about to appear
\cite{Athron:2009bs}.

In section 2 we briefly review the E$_6$SSM, then in section 3 we
introduce the cE$_6$SSM. Section 4 describes the experimental and
theoretical constraints and section 5 discusses the aforementioned
predictions of the cE$_6$SSM elucidated by five ``early discovery''
benchmark points.  Section 6 concludes the Letter.

\section{The E$_6$SSM}

One of the most important issues in models with additional Abelian
gauge symmetries is the cancellation of anomalies. In $E_6$ theories,
if the surviving Abelian gauge group factor is a subgroup of $E_6$,
and the low energy spectrum constitutes a complete $27$ representation
of $E_6$, then the anomalies are cancelled automatically.  The $27_i$
of $E_6$, each containing a quark and lepton family, decompose
under the $SU(5)\times U(1)_{N}$ subgroup of $E_6$ as follows:
\be
27_i\to \ds\left(10,\,\ds{1}\right)_i+\left(5^{*},\,\ds{2}\right)_i
+\left(5^{*},\,-\ds{3}\right)_i +\ds\left(5,-\ds{2}\right)_i
+\left(1,\ds{5}\right)_i+\left(1,0\right)_i\,.
\label{4}
\ee
The first and second quantities in the brackets are the $SU(5)$
representation and extra $U(1)_{N}$ charge while $i$ is a family index
that runs from 1 to 3. From Eq.~(\ref{4}) we see that, in order to
cancel anomalies, the low energy (TeV scale) spectrum must contain
three extra copies of $5^*+5$ of $SU(5)$ in addition to the three
quark and lepton families in $5^*+10$. To be precise, the ordinary SM
families which contain the doublets of left-handed quarks $Q_i$ and
leptons $L_i$, right-handed up- and down-quarks ($u^c_i$ and $d^c_i$)
as well as right-handed charged leptons, are assigned to
$\left(10,\ds{1}\right)_i+\left(5^{*},\,\ds{2}\right)_i$.
Right-handed neutrinos $N^c_i$ should be associated with the last term
in Eq.~(\ref{4}), $\left(1,0\right)_i$.  The next-to-last term in
Eq.~(\ref{4}), $\left(1,\ds{5}\right)_i$, represents SM-type singlet
fields $S_i$ which carry non-zero $U(1)_{N}$ charges and therefore
survive down to the EW scale.  The three pairs of $SU(2)$-doublets
($H^d_{i}$ and $H^u_{i}$) that are contained in
$\left(5^{*},\,-\ds{3}\right)_i$ and $\left(5,-\ds{2}\right)_i$ have
the quantum numbers of Higgs doublets, and we shall identify one of
these pairs with the usual MSSM Higgs doublets, with the other two
pairs being ``inert'' Higgs doublets which do not get VEVs. The other
components of these $SU(5)$ multiplets form colour triplets of exotic
quarks $D_i$ and $\overline{D_i}$ with electric charges $-1/3$ and
$+1/3$ respectively. The matter content and correctly normalised
Abelian charge assignment are in Tab.~\ref{charges}.

\begin{table}[ht]
  \centering
  \begin{tabular}{|c|c|c|c|c|c|c|c|c|c|c|c|c|c|}
    \hline
 & $Q$ & $u^c$ & $d^c$ & $L$ & $e^c$ & $N^c$ & $S$ & $H_2$ & $H_1$ & $D$ &
 $\overline{D}$ & $H'$ & $\overline{H'}$ \\
 \hline
$\sqrt{\frac{5}{3}}Q^{Y}_i$
 & $\frac{1}{6}$ & $-\frac{2}{3}$ & $\frac{1}{3}$ & $-\frac{1}{2}$
& $1$ & $0$ & $0$ & $\frac{1}{2}$ & $-\frac{1}{2}$ & $-\frac{1}{3}$ &
 $\frac{1}{3}$ & $-\frac{1}{2}$ & $\frac{1}{2}$ \\
 \hline
$\sqrt{{40}}Q^{N}_i$
 & $1$ & $1$ & $2$ & $2$ & $1$ & $0$ & $5$ & $-2$ & $-3$ & $-2$ &
 $-3$ & $2$ & $-2$ \\
 \hline
  \end{tabular}
  \caption{\it\small The $U(1)_Y$ and $U(1)_{N}$ charges of matter fields in the
    E$_6$SSM, where $Q^{N}_i$ and $Q^{Y}_i$ are here defined with the correct
$E_6$ normalisation factor required for the RG analysis.}
  \label{charges}
\end{table}

We also require a further pair of superfields $H'$ and $\overline{H}'$
with a mass term $\mu' {H'}{\overline{H}'}$ from incomplete extra
$27'$ and $\overline{27'}$ representations to survive to low energies
to ensure gauge coupling unification. Because $H'$ and $\overline{H}'$
originate from $27'$ and $\overline{27'}$ these supermultiplets do not
spoil anomaly cancellation in the considered model.  Our analysis
reveals that the unification of the gauge couplings in the E$_6$SSM
can be achieved for any phenomenologically acceptable value of
$\alpha_3(M_Z)$, consistent with the measured low energy central
value
\cite{unif-e6ssm}\footnote{The two superfields $H'$ and
$\overline{H}'$ may be removed from the spectrum, thereby avoiding the
$\mu'$ problem, leading to unification at the string scale
\cite{Howl:2008xz}.  However we shall not pursue this possibility in
this Letter.}.

Since right--handed neutrinos have zero charges they can acquire very
heavy Majorana masses. The heavy Majorana right-handed neutrinos may
decay into final states with lepton number $L=\pm 1$, thereby creating
a lepton asymmetry in the early Universe.  Because the Yukawa
couplings of exotic particles are not constrained by the neutrino
oscillation data, substantial values of CP--violating lepton
asymmetries can be induced even for a relatively small mass of the
lightest right--handed neutrino ($M_1 \sim 10^6\,\mbox{GeV}$) so that
successful thermal leptogenesis may be achieved without encountering
any gravitino problem \cite{King:2008qb}.

The superpotential of the E$_6$SSM involves a lot of new Yukawa
couplings in comparison to the SM.  In general these new interactions
violate baryon number conservation and induce non-diagonal flavour
transitions. To suppress baryon number violating and flavour changing
processes one can postulate a $Z^{H}_2$ symmetry under which all
superfields except one pair of $H^d_{i}$ and $H^u_{i}$ (say $H_d\equiv
H^d_{3}$ and $H_u\equiv H^u_{3}$) and one SM-type singlet field
($S\equiv S_3$) are odd.  The $Z^{H}_2$ symmetry reduces the structure
of the Yukawa interactions, and an assumed hierarchical structure of
the Yukawa interactions allows to simplify the form of the E$_6$SSM
superpotential substantially. Keeping only Yukawa interactions whose
couplings are allowed to be of order unity leaves us with
\beq
\ba{rcl}
W_{\rm E_6SSM}&\simeq &\lambda S (H_{d} H_{u})+\lambda_{\alpha}
S(H^d_{\alpha} H^u_{\alpha})+ \kappa_i S (D_i\overline{D}_i)\\[2mm]
&&+h_t(H_{u}Q)t^c+h_b(H_{d}Q)b^c+ h_{\tau}(H_{d}L)\tau^c+
\mu'(H^{'}\overline{H^{'}}),
\ea
\label{cessm8}
\eeq
where $\alpha,\beta=1,2$ and $i=1,2,3$, and where the superfields
$L=L_3$, $Q=Q_3$, $t^c=u^c_3$, $b^c=d^c_3$ and $\tau^c=e^c_3$ belong
to the third generation and $\lambda_i$, $\kappa_i$ are dimensionless
Yukawa couplings with $\lambda \equiv \lambda_3$.  Here we assume that
all right--handed neutrinos are relatively heavy so that they can be
integrated out\footnote{We shall ignore the presence of right-handed
neutrinos in the subsequent RG analysis.}.  The $SU(2)_L$ doublets
$H_u$ and $H_d$, which are even under the $Z^{H}_2$ symmetry, play the
role of Higgs fields generating the masses of quarks and leptons after
EWSB. The singlet field $S$ must also acquire a large VEV to induce
sufficiently large masses for the $Z'$ boson. The couplings
$\lambda_i$ and $\kappa_i$ should be large enough to ensure the exotic
fermions are sufficiently heavy to avoiding conflict with direct
particle searches at present and former accelerators. They should also
be large enough so that the evolution of the soft scalar mass $m_S^2$
of the singlet field $S$ results in negative values of $m_S^2$ at low
energies, triggering the breakdown of the $U(1)_{N}$ symmetry.

However the $Z^{H}_2$ can only be approximate (otherwise the exotics
would not be able to decay).  To prevent rapid proton decay in the
E$_6$SSM a generalised definition of $R$--parity should be used. We
give two examples of possible symmetries that can achieve that.  If
$H^d_{i}$, $H^u_{i}$, $S_i$, $D_i$, $\overline{D}_i$ and the quark
superfields ($Q_i$, $u^c_i$, $d^c_i$) are even under a discrete
$Z^L_2$ symmetry while the lepton superfields ($L_i$, $e^c_i$,
$N^c_i$) are odd (Model I) then the allowed superpotential is
invariant with respect to a $U(1)_B$ global symmetry. The exotic
$\overline{D_i}$ and $D_i$ are then identified as diquark and
anti-diquark, i.e. $B_{D}=-2/3$ and $B_{\overline{D}}=2/3$. An
alternative possibility is to assume that the exotic quarks $D_i$ and
$\overline{D_i}$ as well as lepton superfields are all odd under
$Z^B_2$ whereas the others remain even. In this case (Model II) the
$\overline{D_i}$ and $D_i$ are leptoquarks \cite{King:2005jy}.

After the breakdown of the gauge symmetry, $H_u$, $H_d$ and $S$ form
three CP--even, one CP-odd and two charged states in the Higgs
spectrum.  The mass of one CP--even Higgs particle is always very
close to the $Z'$ boson mass $M_{Z'}$. The masses of another CP--even,
the CP--odd and the charged Higgs states are almost degenerate.
Furthermore, like in the MSSM and NMSSM, one of the CP--even Higgs
bosons is always light irrespective of the SUSY breaking scale.
However, in contrast with the MSSM, the lightest Higgs boson in the
E$_6$SSM can be heavier than $110-120\,\mbox{GeV}$ even at tree
level. In the two--loop approximation the lightest Higgs boson mass
does not exceed $150-155\,\mbox{GeV}$ \cite{King:2005jy}. Thus the
SM--like Higgs boson in the E$_6$SSM can be considerably heavier than
in the MSSM and NMSSM, since it contains a similar F-term contribution
as the NMSSM but with a larger maximum value for $\lambda(m_t)$ as it
is not bounded as strongly by the validity of perturbation theory up to
the GUT scale \cite{King:2005jy}.  However in the considered ``early
discovery'' benchmark points in this Letter, it will always be close
to the current LEP2 limit.

\section{The Constrained E$_6$SSM}

The simplified superpotential of the E$_6$SSM involves seven extra
couplings ($\mu'$, $\kappa_i$ and $\lambda_i$) as compared with the
MSSM with $\mu=0$. The soft breakdown of SUSY gives rise to many new
parameters.  The number of fundamental parameters can be reduced
drastically though within the constrained version of the E$_6$SSM
(cE$_6$SSM).  Constrained SUSY models imply that all soft scalar
masses are set to be equal to $m_0$ at some high energy scale $M_X$,
taken here to be equal to the GUT scale, all gaugino masses $M_i(M_X)$
are equal to $M_{1/2}$ and trilinear scalar couplings are such that
$A_i(M_X)=A_0$. Thus the cE$_6$SSM is characterised by the following
set of Yukawa couplings, which are allowed to be of the order of
unity, and universal soft SUSY breaking terms,
\begin{equation}
\lambda_i(M_X),\quad \kappa_i(M_X),\quad h_t(M_X),\quad h_b(M_X), \quad h_{\tau}(M_X), \quad m_0, \quad M_{1/2},\quad A_0,
\label{3}
\end{equation}
where $h_t(M_X)$, $h_b(M_X)$ and $h_{\tau}(M_X)$ are the usual
$t$--quark, $b$--quark and $\tau$--lepton Yukawa couplings, and
$\lambda_i(M_X)$, $\kappa_i(M_X)$ are the extra Yukawa couplings
defined in Eq.~(\ref{cessm8}). The universal soft scalar and trilinear
masses correspond to an assumed high energy soft SUSY breaking
potential of the universal form,
\begin{equation}
V_{soft}= m_0^227_i27_i^*+A_0Y_{ijk}27_i27_j27_k +h.c.,
\label{potential}
\end{equation}
where $Y_{ijk}$ are generic Yukawa couplings from the trilinear terms
in Eq.~(\ref{cessm8}) and the $27_i$ represent generic fields from
Eq.~(\ref{4}), and in particular those which appear in
Eq.~(\ref{cessm8}). Since $Z^{H}_2$ symmetry forbids many terms in
the superpotential of the E$_6$SSM it also forbids similar soft
SUSY breaking terms in Eq.~(\ref{potential}).
To simplify our analysis we assume that all
parameters in Eq.~(\ref{3}) are real and $M_{1/2}$ is positive.  In
order to guarantee correct EWSB $m_0^2$ has to be positive.  The set
of cE$_6$SSM parameters in Eq.~(\ref{3}) should in principle be
supplemented by $\mu'$ and the associated bilinear scalar coupling
$B'$. However, since $\mu'$ is not constrained by the EWSB and the
term $\mu'H'\overline{H}'$ in the superpotential is not suppressed by
$E_6$, the parameter $\mu'$ will be assumed to be $\sim
10\,\mbox{TeV}$ so that $H'$ and $\overline{H}'$ decouple from the
rest of the particle spectrum. As a consequence the parameters $B'$
and $\mu'$ are irrelevant for our analysis.

To calculate the particle spectrum within the cE$_6$SSM one must find
sets of parameters which are consistent with both the high scale
universality constraints and the low scale EWSB constraints.  To
evolve between these two scales we use two--loop renormalisation group
equations (RGEs) for the gauge and Yukawa couplings together with
two--loop RGEs for $M_a(Q)$ and $A_i(Q)$ as well as one--loop RGEs for
$m_i^2(Q)$. $Q$ is the renormalisation scale. The RGE evolution is
performed using a modified version of SOFTSUSY 2.0.5
\cite{Allanach:2001kg} and the RGEs for the E$_6$SSM are presented
in a longer paper \cite{Athron:2009bs}. The details of the procedure we followed
are summarized below.\\[-4mm]

\noindent 1. The gauge and Yukawa couplings are determined
independently of the soft SUSY breaking mass parameters as follows:

(i) We choose input values for $s = \sqrt{2}\langle S \rangle$ and
$\tan\beta=v_2/v_1$ (where $v_2$ and $v_1$ are the usual VEVs of the
Higgs fields $H_u$ and $H_d$) as defined by our scenario.

(ii) We set the gauge couplings $g_1$, $g_2$ and $g_3$ equal to the
experimentally measured values at $M_Z$.

(iii) We fix the low energy Yukawa couplings $h_t$, $h_b$, and
$h_{\tau}$ using the relations between the running masses of the
fermions of the third generation and VEVs of the Higgs fields, i.e.
\be
m_t(M_t)=\dfrac{h_t(M_t) v}{\sqrt{2}}\sin\beta,~~
m_b(M_t)=\dfrac{h_b(M_t) v}{\sqrt{2}}\cos\beta,~~
m_{\tau}(M_t)=\dfrac{h_{\tau}(M_t) v}{\sqrt{2}}\cos\beta .
\label{cessm29}
\ee

(iv) The gauge and Yukawa couplings are then evolved up to the GUT
scale $M_X$. Using the beta functions for QED and QCD, the gauge
couplings are first evolved up to $m_t$. Since we are employing
two--loop RGEs in the SUSY preserving sector, we include one estimated
threshold scale for the masses of the superpartners of the SM
particles, $T_{MSSM}$, and one for the masses of the new exotic
particles, $T_{ESSM}$.  Since these are common mass scales we neglect
mass splitting within each group of particles. So between $m_t$ and
$T_{MSSM}$ we evolve these gauge and Yukawa couplings with SM RGEs and
between $T_{MSSM}$ and $T_{ESSM}$ we employ the MSSM RGEs. At
$T_{ESSM}$ the values of E$_6$SSM gauge and Yukawa couplings, $g_1$,
$g_2$, $g_3$, $h_t$, $h_b$ and $h_\tau$, form a low energy boundary
condition for what follows. Initial low energy estimates of the new
E$_6$SSM Yukawa couplings, $\lambda_i$ and $\kappa_i$ are also input
here, and all SUSY preserving couplings are evolved up to the high
scale using the two--loop E$_6$SSM RGEs.

(v) At the GUT scale $M_X$ we set
$g_1(M_X)=g_2(M_X)=g_3(M_X)=g_1'(M_X) \equiv g_0$ and select values
for $\kappa_i(M_X)$ and $\lambda_i(M_X)$, which are input parameters
in our procedure. An iteration is then performed between $M_X$ and the
low energy scale to obtain the values of all the gauge and Yukawa
couplings which are consistent with our input values for
$\kappa_i(M_X)$, $\lambda_i(M_X)$, gauge coupling unification and our
low scale boundary conditions, derived from experimental data.\\[-4mm]

\noindent 2. Having completely determined the gauge and Yukawa couplings,
the low energy soft SUSY breaking parameters are then determined
semi-analytically as functions of the GUT scale values of $A_0$,
$M_{1/2}$ and $m_0$. They take the form,
\begin{eqnarray}
m_i^2(Q) &=& a_i(Q)  M_{1/2}^2 + b_i(Q) A_0^2 + c_i(Q) A_0 M_{1/2} + d_i(Q) m_0^2,\\
A_i(Q) &=& e_i(Q) A_0 + f_i(Q) M_{1/2},\\
M_i(Q) &=& p_i(Q) A_0 + q_i(Q) M_{1/2},
\end{eqnarray}
where $Q$ is the renormalisation scale. The coefficients are unknown
but may be determined numerically at the low energy scale, as follows:

(i) Set $A_0=M_{1/2}=0$ at $M_X$ with $m_0$ non-zero, and run the full
set of E$_6$SSM parameters down to the low scale to yield the
coefficients proportional to $m_0^2$ in the expressions for the soft
SUSY breaking parameters.

(ii) Repeat for $A_0$ and $M_{1/2}$.

(iii) The coefficient of the $A_0M_{1/2}$ term is determined using
non-zero values of both $A_0$ and $M_{1/2}$ at $M_X$, using the
results in part (ii) to isolate this term.\\[-4mm]

\noindent 3. Using the semi-analytic expressions for the soft masses
from step 2 above, we then impose conditions for correct EWSB at low
energy and determine sets of $m_0$, $M_{1/2}$ and $A_0$ which are
consistent with EWSB, as follows:

(i) Working with the tree--level potential $V_0$ (to start with) we
impose the minimisation conditions $\dfrac{\partial V_0}{\partial s}=
\dfrac{\partial V_0}{\partial v_1}=\dfrac{\partial V_0}{\partial
v_2}=0$ leading to a system of quadratic equations in $m_0$, $M_{1/2}$
and $A_0$. In this approximation, the equations can be reduced to two
second order equations with respect to $A_0$ and $M_{1/2}$ which can
have up to four solutions for each set of Yukawa couplings.

(ii) For each solution $m_0$, $M_{1/2}$ and $A_0$, the low energy stop
soft mass parameters are determined and the one--loop Coleman-Weinberg
Higgs effective potential $V_1$ is calculated.  The new minimisation
conditions for $V_1$ are then imposed, and new solutions for $m_0$,
$M_{1/2}$ and $A_0$ are obtained.

(iii) The procedure in (ii) is then iterated until we find stable
solutions. Some or all of the obtained solutions can be complex.
Here we restrict our consideration to the scenarios with real
values of fundamental parameters which do not induce any
CP--violating effects. For some values of $\tan\beta$, $s$ and
Yukawa couplings the solutions with real $A_0$, $M_{1/2}$ and
$m_0$ do not exist. There is a substantial part of the parameter
space where there are only two solutions with real values of
fundamental parameters. However there are also some regions of the
parameters where all four solutions of the non--linear algebraic
equations are real.

Although correct EWSB is not guaranteed in the cE$_6$SSM,
remarkably, there are always solutions with real $A_0$, $M_{1/2}$
and $m_0$ for sufficiently large values of $\kappa_i$, which drive
$m_S^2$ negative. This is easy to understand since the $\kappa_i$
couple the singlet to a large multiplicity of coloured fields,
thereby efficiently driving its squared mass negative to trigger
the breakdown of the gauge symmetry.\\[-4mm]

\noindent 4. Using the obtained solutions we calculate the masses of all
exotic and SUSY particles for each set of fundamental parameters
in Eq.~(\ref{3}).\\[-4mm]

Finally at the last stage of our analysis we vary Yukawa couplings,
$\tan\beta$ and $s$ to establish the qualitative pattern of the
particle spectrum within the cE$_6$SSM. To avoid any conflict with
present and former collider experiments as well as with recent
cosmological observations we impose the set of constraints specified
in the next section. These bounds restrict the allowed range of the
parameter space in the cE$_6$SSM.

\section{Experimental and Theoretical Constraints}

The experimental constraints applied in our analysis are: $m_h \geq
114$ GeV, all sleptons and charginos are heavier than
$100\,\mbox{GeV}$, all squarks and gluinos have masses above
$300\,\mbox{GeV}$ and the $Z'$ boson has a mass which is larger than
$861\,\mbox{GeV}$ \cite{cdfZZ}. We also impose the most conservative
bound on the masses of exotic quarks and squarks that comes from the
HERA experiments \cite{Aktas:2005pr}, by requiring them to be heavier
than $300\,\mbox{GeV}$. Finally we require that the inert Higgs and
inert Higgsinos are heavier than 100 GeV to evade limits on Higgsinos
and charged Higgs bosons from LEP.

In addition to a set of bounds coming from the non--observation of new
particles in experiments we impose a few theoretical constraints. We
require that the lightest SUSY particle (LSP) should be a
neutralino. We also restrict our consideration to values of the Yukawa
couplings $\lambda_i(M_X)$, $\kappa_i(M_X)$, $h_t(M_X)$, $h_b(M_X)$
and $h_{\tau}(M_X)$ less than 3 to ensure the applicability of
perturbation theory up to the GUT scale.

\begin{figure}[h]
\begin{center}
\resizebox{!}{10cm}{%
\includegraphics{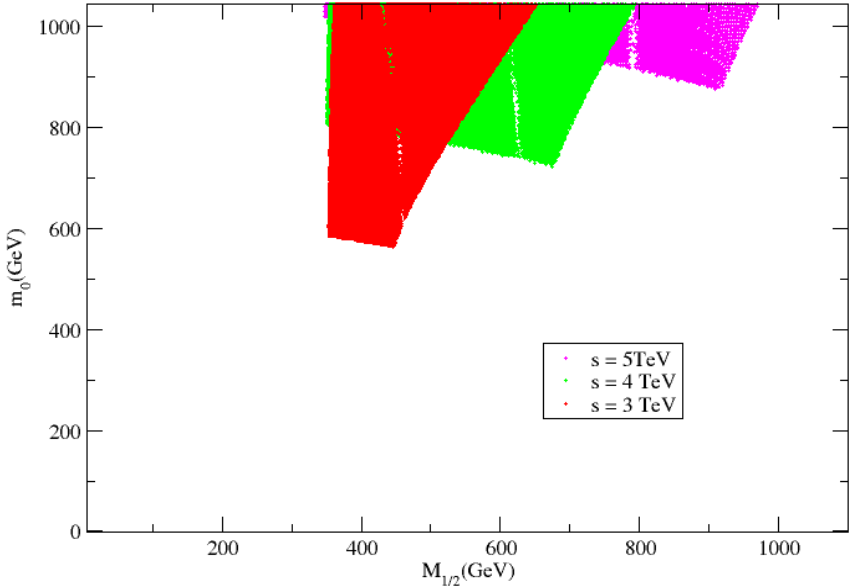}
}
\caption{Physical solutions with $\tan \beta = 10$, $\lambda_{1,2} =
0.1$, $s= 3, 4, 5$ TeV fixed and $\lambda \equiv \lambda_3$ and
$\kappa \equiv \kappa_{1,2,3}$ varying, which pass experimental
constraints from LEP and Tevatron data.  On the left-hand side of each
allowed region the chargino mass is less than $100$ GeV, while
underneath the inert Higgses are less than $100$ GeV or becoming
tachyonic. The region ruled out immediately to the right of the
allowed points is due to $m_h < 114 $ GeV.
The results show that $m_0>M_{1/2}$ for each value of
$s$. They also show that higher $M_{1/2}$ are correlated with
higher $s$ (and thus higher $Z'$ masses).
\label{tb10_s3TeV_Valid} }
\end{center}

\end{figure}

In our exploration of the cE$_6$SSM parameter space we first looked at
scenarios with a universal coupling between exotic coloured
superfields and the third generation singlet field ${S}$, $\kappa(M_X)
= \kappa_{1,2,3}(M_X)$, and fixed the inert Higgs couplings
$\lambda_{1,2}(M_X) = 0.1$. In fixing $\lambda_{1,2}$ like this we are
deliberately pre-selecting for relatively light inert Higgsinos. The
third generation Yukawa $\lambda = \lambda_3$ was allowed to vary
along with $\kappa$.  Splitting $\lambda_3$ from $\lambda_{1,2}$ seems
reasonable since $\lambda_3$ plays a very special role in E$_6$SSM
models in forming the effective $\mu$-term when $S$ develops a VEV.
Eventually we allowed non--universal $\kappa_i(M_X)$.  For fixed
values of $\tan \beta = 3, 10, 30$, we scanned over $s, \kappa,
\lambda$.  From these input parameters, the sets of soft mass
parameters, $A_0$, $M_{1/2}$ and $m_0$, which are consistent with the
correct breakdown of electroweak symmetry, are found.  We find that
for fixed values of the Yukawas the soft mass parameters scale with
$s$, while if $s$ and $\tan \beta$ are fixed, varying the Yukawas,
$\lambda$ and $\kappa$, then produces a bounded region of allowed
points.  The value of $s$ determines the location and extent of the
bounded regions.  As $s$ is increased the lowest values of $m_0$ and
$M_{1/2}$, consistent with experimental searches and EWSB
requirements, increase. This is shown in Fig.~\ref{tb10_s3TeV_Valid}
where the allowed regions for three different values of the singlet
VEV, $s = 3$, $4$ and $5$ TeV, are compared, with the allowed regions
in red, green, magenta respectively and the excluded regions in white.
These regions overlap since we are finding soft masses consistent with
EWSB conditions that have a non--linear dependence on the VEVs and
Yukawas.

\section{Predictions of the cE$_6$SSM}

\subsection{Overview of the spectrum and decay signatures}
\subsubsection{SUSY spectrum and signatures}
From Fig.~\ref{tb10_s3TeV_Valid} we see that $m_0>M_{1/2}$ for each
value of $s$ and also that lower $M_{1/2}$ is weakly correlated with
lower $s$ and thus lower $Z'$ masses. As is discussed in detail in
Ref.~\cite{Athron:2009bs} this bound is caused, depending on the value
of $\tan \beta$, either by the inert Higgs masses being driven below
their experimental limit from negative D-term contributions canceling
the positive contribution from $m_0$ or the light Higgs mass going
below the LEP2 limit. 

Another remarkable feature of the cE$_6$SSM is that the
low energy gluino mass parameter $M_3$ is driven to be smaller than
$M_{1/2}$ by RG running.  The reason for this is that the E$_6$SSM has
a much larger (super)field content than the MSSM (three 27's instead
of three 16's), so much so that at one--loop order the QCD beta
function (accidentally) vanishes in the E$_6$SSM, and at two loops it
loses asymptotic freedom (though the gauge couplings remain
perturbative at high energy). This implies that the low energy gaugino
masses are all less than $M_{1/2}$ in the cE$_6$SSM, being given by
roughly $M_3 \sim 0.7M_{1/2}$, $M_2 \sim 0.25M_{1/2}$, $M_1 \sim
0.15M_{1/2}$. These should be compared to the corresponding low energy
values in the MSSM, $M_3 \sim 2.7M_{1/2}$, $M_2 \sim 0.8M_{1/2}$, $M_1
\sim 0.4M_{1/2}$.  Thus, in the cE$_6$SSM, since the low energy
gaugino masses $M_i$ are driven by RG running to be small, the
lightest SUSY states will generally consist of a light gluino of mass
$\sim M_3$, a light wino-like neutralino and chargino pair of mass
$\sim M_2$, and a light bino-like neutralino of mass $\sim M_1$, which
are typically all much lighter than the Higgsino masses of order $\mu
= \lambda s/\sqrt{2}$, where $\lambda$ cannot be too small for correct
EWSB.   Since $m_0>M_{1/2}$ the squarks and sleptons are also
much heavier than the light gauginos.

Thus, throughout all cE$_6$SSM regions of parameter space there is the
striking prediction that the lightest sparticles always include the
gluino $\tilde g$, the two lightest neutralinos $\chi_1^0,\chi_2^0$,
and a light chargino $\chi_1^\pm$.  Therefore pair production of
$\chi_2^0\chi_2^0$, $\chi_2^0\chi_1^\pm$, $\chi_1^\pm \chi_1^\mp$ and
$\tilde g \tilde g$ should always be possible at the LHC irrespective
of the $Z'$ mass.  Due to the hierarchical spectrum, the gluinos can
be relatively narrow states with width $\Gamma_{\tilde{g}}\propto
M_{\tilde{g}}^5/m_{\tilde{q}}^4$, comparable to that of $W^{\pm}$ and
$Z$ bosons.  They will decay through $\tilde{g} \rightarrow q
\tilde{q}^* \rightarrow q \bar{q} + E_T^{\rm miss}$, so gluino pair
production will result in an appreciable enhancement of the cross
section for $pp \rightarrow q \bar q q \bar q + E_T^{\rm miss} + X$,
where $X$ refers to any number of light quark/gluon jets.

The second lightest neutralino decays through $\chi_2^0 \rightarrow
\chi_1^0 + l \bar l $ and so would produce an excess in $pp
\rightarrow l \bar l l \bar l + E_T^{\rm miss} + X$, which could be
observed at the LHC. Since all squarks and sleptons, as well as new
exotic particles, turn out to be rather heavy compared to the low
energy wino mass, the calculation of the branching ratio $Br (\chi_2^0
\rightarrow \chi_1^0 + l \bar l )$ is very similar to that in the
MSSM. This branching ratio in the MSSM is known to be very sensitive
to the choice of fundamental parameters of the model.  For the type of
the neutralino spectra presented later, in which the second lightest
neutralino is approximately wino, the lightest neutralino is
approximately bino, and where the other sparticles are much heavier,
${\rm Br} (\chi_2^0 \rightarrow \chi_1^0 + l \bar l )$ is known to vary from
$1.5\%$ to $6\%$ \cite{neutr-decay}.

\subsubsection{Exotic spectrum and signatures \label{5.1.2}}

Other possible manifestations of the E$_6$SSM at the LHC are related
to the presence of a $Z'$ and exotic multiplets of matter.  The
production of a TeV scale $Z'$ will provide an unmistakable and
spectacular LHC signal even with first data \cite{King:2005jy}.  At
the LHC, the $Z'$ boson that appears in the $E_6$ inspired models can
be discovered if it has a mass below $4-4.5\,\mbox{TeV}$
\cite{ZprimeE6}.  The determination of its couplings should be
possible if $M_{Z'}\lesssim 2-2.5\,\mbox{TeV}$ \cite{Dittmar:2003ir}.

When the Yukawa couplings $\kappa_i$ of the exotic fermions $D_i$ and
$\overline{D}_i$ have a hierarchical structure, some of them can be
relatively light so that their production cross section at the LHC can
be comparable with the cross section of $t\bar{t}$ production
\cite{King:2005jy}. In the E$_6$SSM, the $D_i$ and $\overline{D}_i$
fermions are SUSY particles with negative $R$--parity so they must be
pair produced and decay into quark--squark (if diquarks) or
quark--slepton, squark--lepton (if leptoquarks), leading to final
states containing missing energy from the LSP.

The lifetime and decay modes of the exotic coloured fermions are
determined by the $Z_2^H$ violating couplings.  If $Z_2^H$ is broken
significantly the presence of the light exotic quarks gives rise to a
remarkable signature.  Assuming that $D_i$ and $\overline{D}_i$
fermions couple most strongly to the third family (s)quarks and
(s)leptons, the lightest exotic $D_i$ and $\overline{D}_i$ fermions
decay into $\tilde{t}b$, $t\tilde{b}$, $\bar{\tilde{t}}\bar{b}$,
$\bar{t}\bar{\tilde{b}}$ (if they are diquarks) or $\tilde{t}\tau$,
$t\tilde{\tau}$, $\tilde{b} \nu_{\tau}$, $b\tilde{\nu_{\tau}}$ (if
they are leptoquarks). This can lead to a substantial enhancement of
the cross section of either $pp\to t\bar{t}b\bar{b}+E^{\rm
miss}_{T}+X$ (if diquarks) or $pp\to t\bar{t}\tau \bar{\tau}+E^{\rm
miss}_{T}+X$ or $pp\to b\bar{b}+ E^{\rm miss}_{T}+X$ (if
leptoquarks). Notice that SM production of $ t \bar t \tau^+ \tau ^-$
is $(\alpha_W / \pi)^2$ suppressed in comparison to the leptoquark
decays. Therefore light leptoquarks should produce a strong signal
with low SM background at the LHC.  In principle the detailed LHC
analyses is required to establish the feasibility of extracting the
excess of $t\bar{t}b\bar{b}$ or $t\bar{t}\tau^+\tau ^-$ production
induced by the light exotic quarks predicted by our model.

We have already remarked that the lifetime and decay modes of the
exotic coloured fermions are determined by the $Z_2^H$ violating
couplings. If $Z_2^{H}$ is only very slightly broken exotic quarks may
be very long lived, with lifetimes up to about 1 s.  This is the
case, for example, in some minimal versions of the model
\cite{Howl:2008xz}.  In this case the exotic $D_i$ and
$\overline{D}_i$ fermions could hadronize before decaying, leading to
spectacular signatures consisting of two low multiplicity jets, each
containing a single quasi-stable heavy D-hadron, which could be
stopped for example in the muon chambers, before decaying much later.

In Tab.~\ref{Xsec} we estimate the total production cross section of
exotic quarks at the LHC for a few different values of their masses
assuming that all masses of exotic quarks are equal
(i.e. $\mu_{D_{i}}=\mu_D$) and all sparticles as well as other new
exotic particles are very heavy. The results in Tab.~\ref{Xsec}
suggest that the observation of the $D$ fermions might be possible if
they have masses below about 1.5-2 TeV \cite{King:2005jy}.

\begin{table}[ht]
\centering
\begin{tabular}{|c|c|c|c|c|c|c|c|c|}
\hline
$\mu_D$[GeV] &
300 & 400 & 500 & 700 & 1000 & 1500 & 2000 & 3000 \\
\hline
$\sigma(pp\to D\bar{D})$[pb] &
76.4 & 17.4 & 5.30 & 0.797 & 0.0889 & $4.94\cdot 10^{-3}$ &
$4.09\cdot 10^{-4}$ & $3.51\cdot 10^{-6}$ \\
\hline
\end{tabular}
\caption{\it\small The cross section of $D\bar{D}$ production at the
LHC as a function of the masses of exotic quarks. For simplicity we
assume that three families of exotic quarks have the same masses.}
\label{Xsec}
\end{table}

Similar considerations apply to the case of exotic $\tilde{D}_i$ and
$\tilde{\overline{D}}_i$ scalars except that they are non--SUSY
particles so they may be produced singly and decay into quark--quark
(if diquarks) or quark--lepton (if leptoquarks) without missing energy
from the LSP.  It is possible to have relatively light exotic coloured
scalars due to mixing effects. The RGEs for the soft breaking masses,
$m_{\tilde{D}_i}^2$ and $m_{\tilde{\overline{D}}_i}^2$, are very
similar, with $\frac{d}{dt} (m_{\tilde{D}_i}^2
-m_{\tilde{\overline{D}}_i}^2)=g_1'^2 M_1'^2 $, resulting in
comparatively small splitting between these soft masses. Consequently,
mixing can be large even for moderate values of the $A_0$, leading to
a large mass splitting between the two scalar partners of the exotic
coloured fermions.\footnote{Note that the diagonal entries of the
exotic squark mass matrices have substantial negative contributions
from the $U(1)_N$ D--term quartic interactions in the scalar
potential. These contributions reduce the masses of exotic squarks and
also contribute to their mass splitting since the $U(1)_N$ charges of
$D_i$ and $\overline{D}_i$ are different.} Recent, as yet
unpublished, results from Tevatron searches for dijet resonances
\cite{CDFtevDijet} rule out scalar diquarks with mass less than
$630$ GeV. However, scalar leptoquarks may be as light as $300$ GeV
since at hadron colliders they are pair produced through gluon fusion.
Scalar leptoquarks decay into quark--lepton final states through small
$Z_2^H$ violating terms, for example $\tilde D \rightarrow t \tau$,
and pair production leads to an enhancement of $pp \rightarrow t \bar
t \tau \bar{\tau}$ (without missing energy) at the LHC.

In addition, the inert Higgs bosons and Higgsinos (i.e.\ the first and
second families of Higgs doublets predicted by the E$_6$SSM which
couple weakly to quarks and leptons and do not get VEVs) can be light
or heavy depending on their free parameters.  The light inert Higgs
bosons decay via $Z_2^H$ violating terms which are analogous to the
Yukawa interactions of the Higgs superfields, ${H}_u$ and $H_d$. One
can expect that the couplings of the inert Higgs fields would have a 
similar hierarchical structure as the couplings of the normal Higgs
multiplets, therefore we assume the $Z_2^H$ breaking interactions
predominantly couple the inert Higgs bosons to the third generation.
So the neutral inert Higgs bosons decay predominantly into 3rd
generation fermion--anti-fermion pairs like $H^{0}_{1,\,i} \rightarrow
b \bar b$. The charged inert Higgs bosons also decay into
fermion--anti-fermion pairs, but in this case it is the antiparticle
of the fermions' EW partner, e.g.~$H^{-}_{1,\,i} \rightarrow \tau
\bar{\nu}_{\tau}$.  The inert Higgs bosons may also be quite heavy, so
that the only light exotic particles are the inert Higgsinos.  Similar
couplings govern the decays of the inert Higgsinos; the
electromagnetically neutral Higgsinos predominantly decay into
fermion-anti-sfermion pairs (e.g.\ $\tilde H_{i}^0 \rightarrow t
\tilde{\bar{t}}^*$, $\tilde H_{i}^0 \rightarrow \tau
\tilde{\bar{\tau}}^*$). The charged Higgsinos decays similarly but in
this case the sfermion is the SUSY partner of the EW partner of the
fermion (e.g.\ $\tilde H_{i}^{+} \rightarrow t \tilde{\bar{b}}^*$,
$\tilde H_{i}^{-} \rightarrow \tau \tilde{\bar{\nu}}_{\tau}^*$).

\subsection{Early discovery benchmarks}
\subsubsection{The benchmark input parameters}
In Tab.~\ref{table:benchmarks} we present a set of ``early discovery''
benchmark points, each associated with a $Z'$ gauge boson mass close
to 1 TeV which should be discovered using first LHC data. The first
block of Tab.~\ref{table:benchmarks} shows the input parameters which
define the benchmark points.  We have selected $s=2.7-3.3$ TeV
corresponding to $M_{Z'}=1020-1250$ GeV, where $M_{Z'}\approx
g'_1sQ_S$ with $Q_S=5/\sqrt{40}$ and $g'_1\approx g_1$.  We have also
restricted ourselves to $(m_0,M_{1/2})<(700,400)$ GeV leading to very
light gauginos, associated with the three low gaugino masses $M_i$,
and in addition a light stop and Higgs mass. Note that for all the
benchmark points the trilinear soft mass is fixed to lie in the range
$A_0=650-1150$ GeV in order to achieve EWSB.

The benchmark points cover three different values of $\tan \beta =
3,10,30$. In each case we have taken $|\lambda_3|$ to be larger than
$\lambda_{1,2}=0.1$ (fixed) at the GUT scale.  In benchmark points A,
B, E (corresponding to $\tan \beta = 3,10,30$) we have taken the
$\kappa_i$ to be universal at the GUT scale and large enough to
trigger EWSB. Since the $\kappa_i$'s control the exotic coloured
fermion masses, this implies that all the $D_i$ and $\overline{D}_i$
fermions are all very heavy in these cases. However it is not
necessary for the $\kappa_i$'s to be universal and these Yukawa
couplings may be hierarchical as for the quark and lepton
couplings. To illustrate this possibility we have considered two
benchmark points, C and D, both for $\tan \beta = 10$, in which
$\kappa_3 \gg \kappa_{1,2}$ at the GUT scale.  In these points C, D we
have taken $\kappa_3$ to be large enough to trigger EWSB, while
allowing $\kappa_{1,2}$ to be low enough to yield light $D_{1,2}$ and
$\overline{D}_{1,2}$ fermion masses.

\subsubsection{The benchmark spectra}
The full spectrum for each of the benchmark points is given in
Tab.~\ref{table:benchmarks} and illustrated in Fig.~\ref{benchmarks}.
The benchmark points all exhibit the characteristic SUSY spectrum
described above of a light gluino $\tilde g$, two light neutralinos
$\chi_1^0,\chi_2^0$, and a light chargino $\chi_1^\pm$. The lightest
neutralino $\chi_1^0$ is essentially pure bino, while $\chi_2^0$ and
$\chi_1^\pm$ are the degenerate components of the wino. Since
$M_{1/2}<400$ GeV for all the points the (two-loop corrected) gluino
mass is below 350 GeV, and the wino mass just above the LEP2 limit of
100 GeV, while the bino is around 60 GeV in each case. The question of
the resulting cosmological dark matter relic abundance is not
considered in this Letter but one should not regard such points with a
light bino as being excluded by cosmology for reasons that will be
discussed later. The Higgsino states are much heavier with the
degenerate Higgsinos $\chi_{3,4}^0$ and $\chi_2^\pm$ having masses
given by $\mu = \lambda s/\sqrt{2}$ in the range 675--830 GeV for all
the benchmark points.  The remaining neutralinos are dominantly
singlet Higgsinos with masses approximately given by $M_{Z'}$.

The Higgs spectrum for all the benchmark points contains a very light
SM--like CP--even Higgs boson $h_1$ with a mass close to the LEP limit
of 115 GeV, making it accessible to LHC or even Tevatron.  The heavier
CP--even Higgs $h_2$, the CP--odd Higgs $A_0$, and the charged Higgs
$H^{\pm}$ are all closely degenerate with masses in the range
600--1000 GeV making them difficult to discover.  The remaining mainly
singlet CP--even Higgs $h_3$ is closely degenerate with the $Z'$.

For benchmarks A, B, E (corresponding to $\tan \beta = 3,10,30$) we
have taken the $\kappa_i$ to be universal and the exotic coloured
fermions have masses in the range 1--1.5 TeV.  However, due to the
mixing effect mentioned previously, we find a light exotic coloured
scalar with a mass of 393 GeV for point E and one at 628 GeV for B.
For benchmark points C and D, with $\kappa_3 \gg \kappa_{1,2}$ at the
GUT scale, there are light exotic coloured fermions in the range
300--400 GeV, together with a light exotic coloured scalar as before.

The inert Higgs masses may be very light depending on the particular
parameters chosen.  For example, for benchmarks B and E the lightest
inert Higgs bosons of the first and second generation have relatively
small masses ($m_{H_{1,\,i}} = 154$ GeV and $m_{H_{1,\,i}} = 220$ GeV
respectively). For all the benchmarks the inert Higgsinos are light,
as $\mu_{\tilde{H}_{i}} = 230-300$ GeV.

The lightest stop mass is in the range $430-550$ GeV for all the
benchmark points, with the remaining squark and slepton masses being
all significantly heavier than the stop mass but below 1 TeV.  Note
that the gluino mass, being below 350 GeV, is always lighter than all
the squark masses for all the benchmark points.

\section{Conclusions}
We have discussed the predictions of a constrained version of the
exceptional supersymmetric standard model (cE$_6$SSM), based on a
universal high energy soft scalar mass $m_0$, soft trilinear mass
$A_0$ and soft gaugino mass $M_{1/2}$. We have seen that the cE$_6$SSM
predicts a characteristic SUSY spectrum containing a light gluino, a
light wino-like neutralino and chargino pair, and a light bino-like
neutralino, with other sparticle masses except the lighter stop being
much heavier.  In addition, the cE$_6$SSM allows the possibility of
light exotic colour triplet charge $1/3$ fermions and scalars, leading
to early exotic physics signals at the LHC.

\begin{figure}[h]
 \begin{center}
\resizebox{!}{6cm}
{\includegraphics{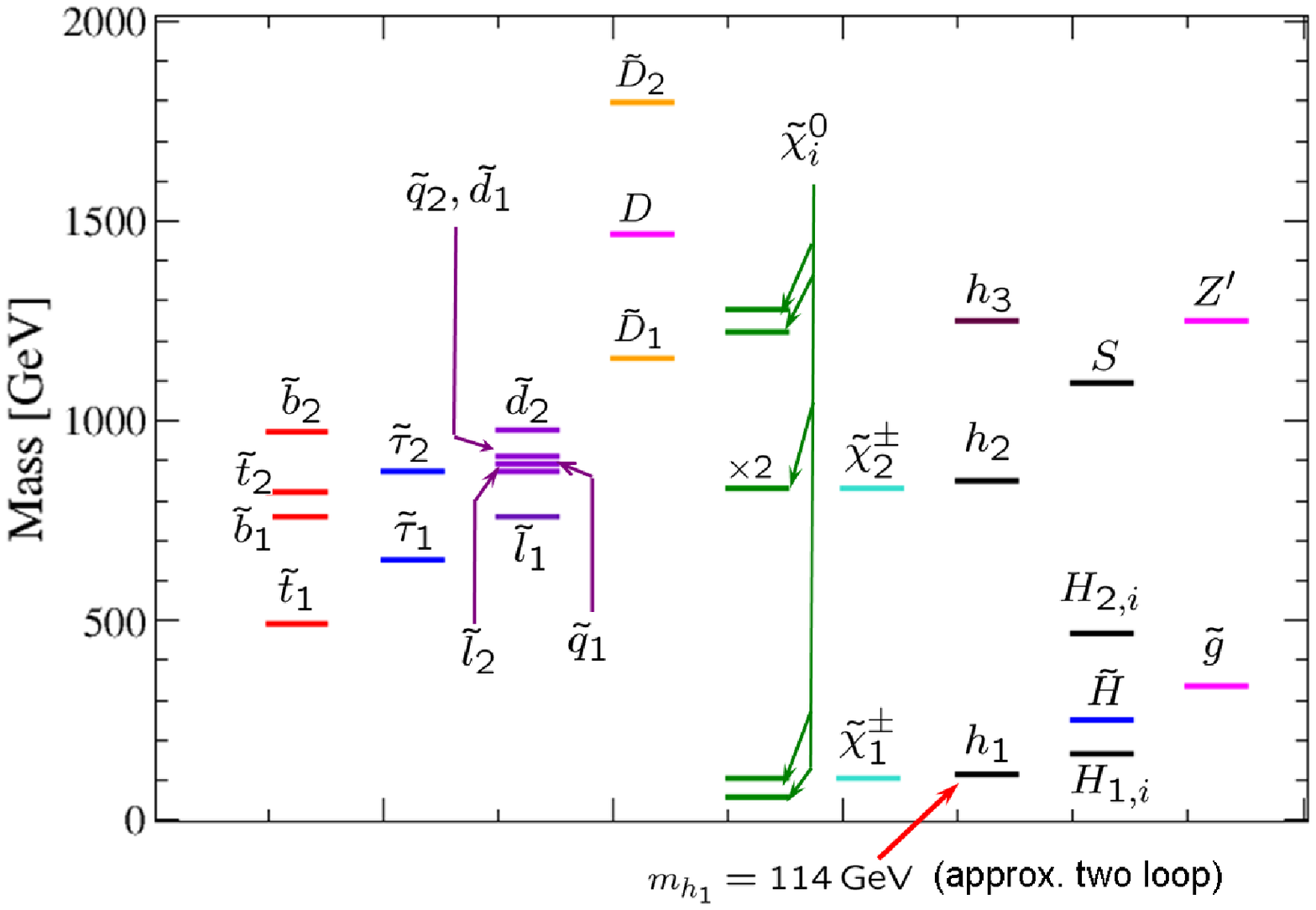}}
\resizebox{!}{6cm}
{\includegraphics{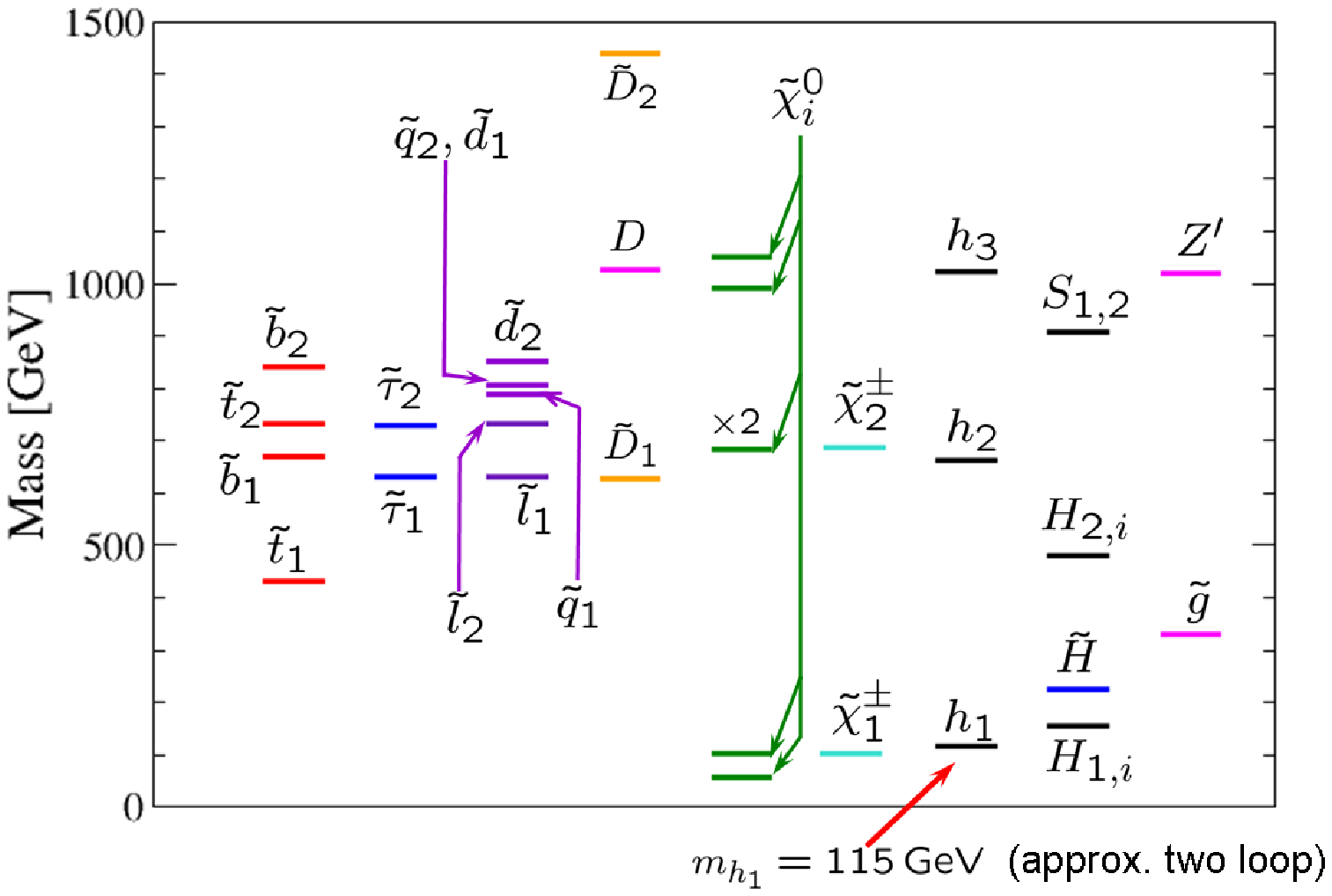}}
\resizebox{!}{6cm}
{\includegraphics{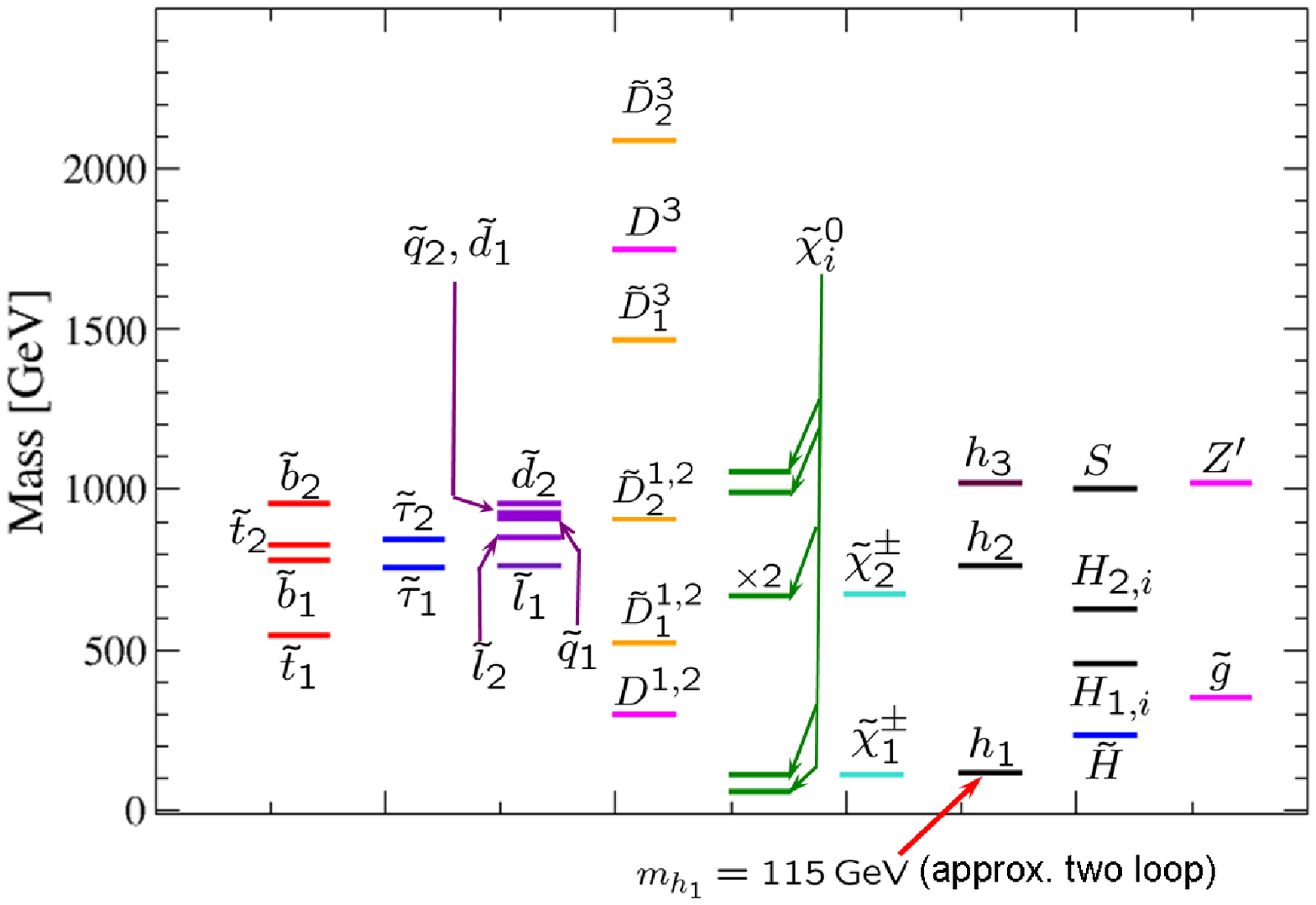}}
\resizebox{!}{6cm}
{\includegraphics{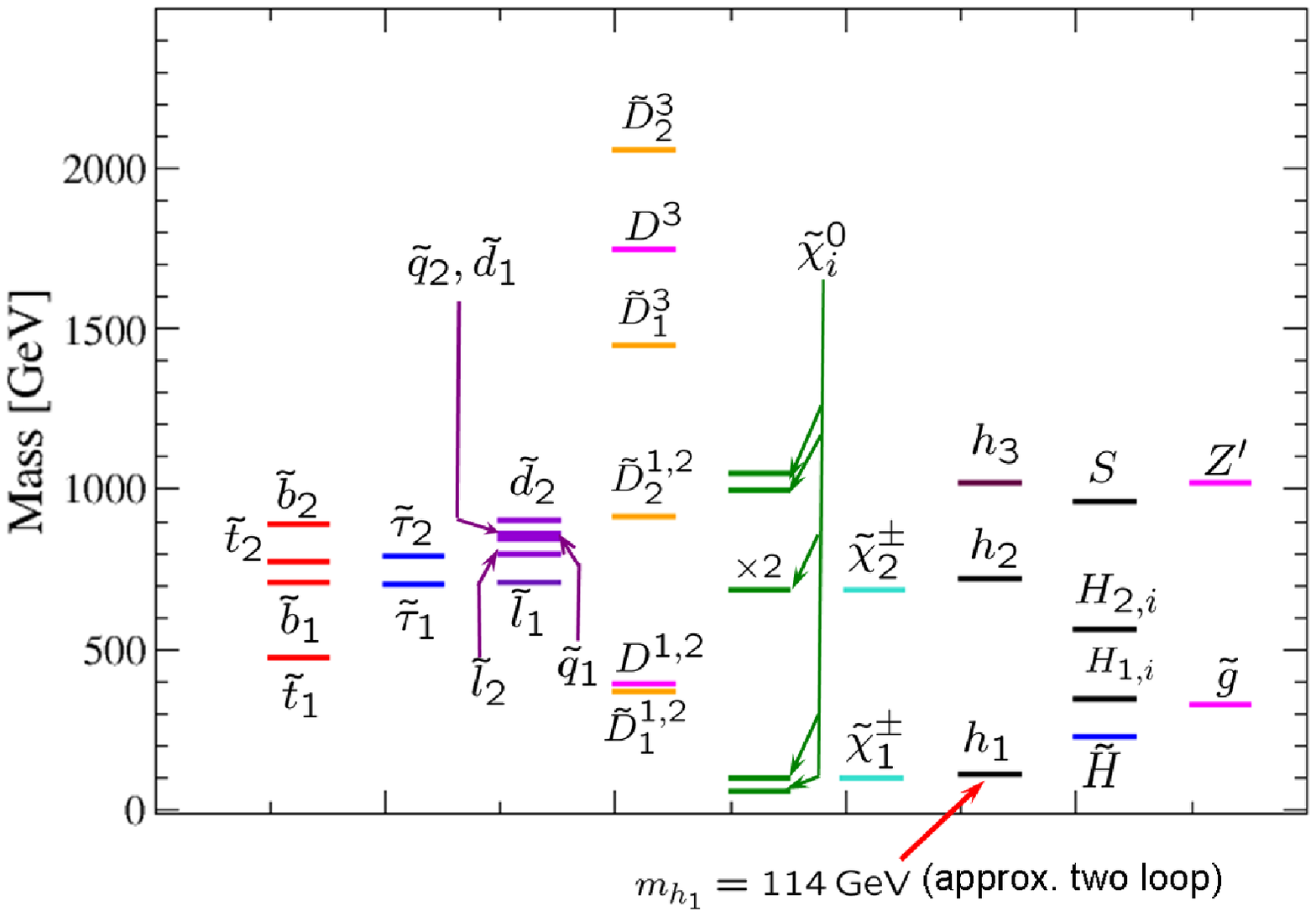}}
\resizebox{!}{6cm}
{\includegraphics{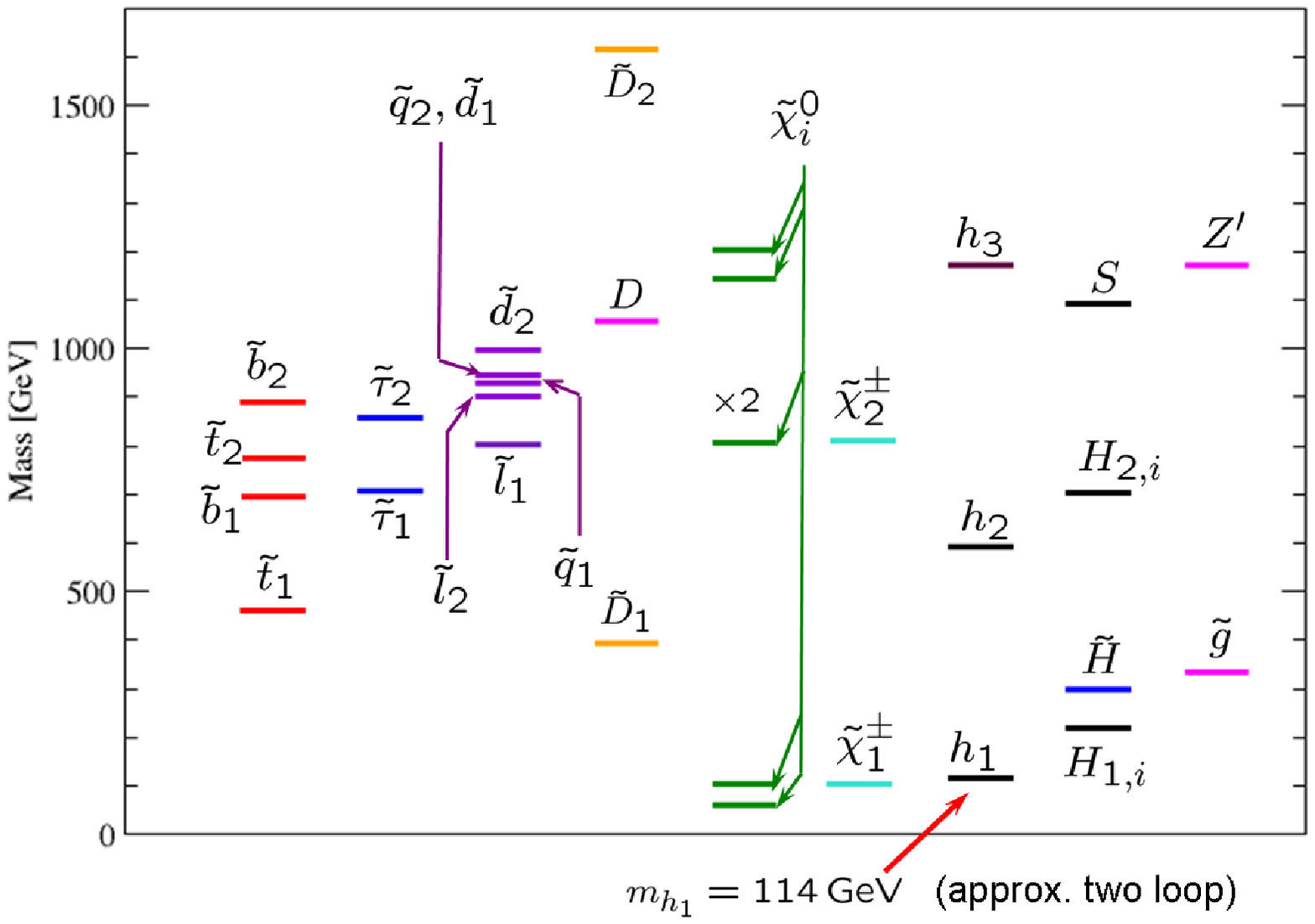}}
\caption{Spectra for
the ``early discovery'' benchmark points A (top left),
B (top right), C (middle left), D (middle right) and
E (bottom centre). \label{benchmarks} }
\end{center}
\end{figure}

We have focussed on the possibility of low values of
$(m_0,M_{1/2})<(700,400)$ GeV, and a $Z'$ gauge boson with mass close
to 1 TeV, which would correspond to an early LHC discovery using
``first data'', and have proposed a set of benchmark points to
illustrate this in Tab.~\ref{table:benchmarks} and
Fig.~\ref{benchmarks}. For some of the benchmarks (C and D) there are
exotic colour triplet charge $\pm 1/3$ $D$ fermions and scalars below
500 GeV, with distinctive final states as discussed in
Section~\ref{5.1.2}. All the benchmark points have a SM--like Higgs
close to the LEP2 limit of 115 GeV with the rest of the Higgs spectrum
significantly heavier. The inert Higgs bosons may be relatively light,
but will be difficult to produce, having zero VEVs and small couplings
to quarks and leptons. The lightest stop mass is in the range
$430-550$ GeV for all the benchmark points, with the remaining squark
and slepton masses being all significantly heavier than the stop mass
but below 1 TeV. The gluino mass is very light, being below 350 GeV in
all cases, and in particular is lighter than the stop squark for all
the benchmark points. The chargino and second neutralino masses are
just above the LEP2 limit of 100 GeV, while the lightest neutralino is
around 60 GeV.

We have not considered the question of cosmological cold dark matter
(CDM) relic abundance due to the neutralino LSP and so one may be
concerned that a bino-like lightest neutralino mass of around 60 GeV
might give too large a contribution to $\Omega_{CDM}$. Indeed a recent
calculation of $\Omega_{CDM}$ in the USSM \cite{Kalinowski:2008iq},
which includes the effect of the MSSM states plus the extra $Z'$ and
the active singlet $S$, together with their superpartners, indicates
that for the benchmarks considered here that $\Omega_{CDM}$ would be
too large.  However the USSM does not include the effect of the extra
inert Higgs and Higgsinos that are present in the E$_6$SSM. While we
have considered the inert Higgsino masses given by
$\mu_{\tilde{H}_{\alpha}} = \lambda_{\alpha} s/\sqrt{2}$, we have not
considered the mass of the inert singlinos. These are generated by
mixing with the Higgs and inert Higgsinos, and are thus of order
$fv^2/s$, where $f$ are additional Yukawa couplings that we have not
specified in our analysis. Since $s\gg v$ it is quite likely that the
LSP neutralino in the cE$_6$SSM will be an inert singlino with a mass
lighter than 60 GeV. This would imply that the state $\chi_1^0$
considered here is not cosmologically stable but would decay into
lighter (essentially inert) singlinos. Such inert singlinos can
annihilate via an s-channel Z-boson, due to their doublet component,
yielding an acceptable CDM relic abundance, as has been recently been
demonstrated in the E$_6$SSM \cite{Hall:2009aj}.  The question of the
calculation of the relic abundance of such an LSP singlino within the
framework of the cE$_6$SSM is beyond the scope of this Letter and will
be considered elsewhere.  In summary, it is clear that one should not
regard the benchmark points with $|m_{\chi^0_1}|\approx 60$ GeV as
being excluded by $\Omega_{CDM}$.

To conclude, in this Letter we have argued that the cE$_6$SSM is a
very well motivated SUSY model and leads to distinctive predictions at
the LHC. We have presented sample benchmark points for which not only
the Higgs boson, but also SUSY particles such as gauginos and stop,
and even more exotic states such as a light $Z'$ and colour triplet
charge $\pm 1/3$ $D$ fermions and scalars, could be just around the
corner in early LHC data. If such states are discovered, this would
not only represent a revolution in particle physics, but would also
point towards an underlying high energy $E_6$ gauge structure,
providing a window into string theory.

\section*{Acknowledgements}
\vspace{0mm} We would like to thank A.~Belyaev, C.~D.~Froggatt and
D.~Sutherland for fruitful discussions. R.N.\ is also grateful to
E.~Boos, M.~I.~Vysotsky and P.~M.~Zerwas for valuable comments
and remarks. D.J.M.\ acknowledges support from the STFC Advanced Fellowship
grant PP/C502722/1. R.N.\ acknowledges support from the SHEFC grant
HR03020 SUPA 36878. S.F.K.\ acknowledges partial
support from the following grants: STFC Rolling Grant
ST/G000557/1 (also SM); EU Network MRTN-CT-2004-503369; NATO grant
PST.CLG.980066 (also SM); EU ILIAS RII3-CT-2004-506222. S.M.\ is also
partially supported by the FP7 RTN
MRTN-CT-2006-035505.

\newpage

\begin{table}[h!]
\begin{center}
\begin{tabular}{|c|c|c|c|c|c|}
\hline  &                \textbf{\footnotesize A} & \textbf{\footnotesize B} & \textbf{\footnotesize C} & \textbf{\footnotesize D} & \textbf{\footnotesize E}   \\\hline
\footnotesize $\tan \beta$             &  \footnotesize  3    &\footnotesize 10     &\footnotesize 10    &\footnotesize 10     & \footnotesize    30    \\[-1.5mm]
\footnotesize $\lambda_3(M_X)$         &\footnotesize   -0.465   &\footnotesize -0.37 &\footnotesize -0.378  &\footnotesize -0.395   &\footnotesize   -0.38\\[-1.5mm]
\footnotesize $\lambda_{1,2}(M_X)$     &\footnotesize    0.1   &\footnotesize 0.1    &\footnotesize 0.1   &\footnotesize 0.1    &\footnotesize    0.1\\[-1.5mm]
\footnotesize $\kappa_3(M_X)$          &\footnotesize    0.3   &\footnotesize 0.2    &\footnotesize 0.42  &\footnotesize 0.43   &\footnotesize    0.17\\[-1.5mm]
\footnotesize $\kappa_{1,2}(M_X)$      &\footnotesize    0.3   &\footnotesize 0.2    &\footnotesize 0.06  &\footnotesize 0.08   &\footnotesize    0.17\\[-1.5mm]
\footnotesize $s$[TeV]                 &\footnotesize    3.3  &\footnotesize 2.7   &\footnotesize 2.7  &\footnotesize 2.7   &\footnotesize    3.1\\[-1.5mm]
\footnotesize $M_{1/2}$[GeV]           &\footnotesize    365   &\footnotesize 363    &\footnotesize 388   &\footnotesize 358    &\footnotesize    365 \\[-1.5mm]
\footnotesize $m_0$ [GeV]              &\footnotesize    640   &\footnotesize 537    &\footnotesize 681   &\footnotesize 623    &\footnotesize    702 \\[-1.5mm]
\footnotesize $A_0$[GeV]                 &\footnotesize    798 &\footnotesize 711    &\footnotesize 645   &\footnotesize 757   &\footnotesize    1148 \\
\hline
\footnotesize $m_{\tilde{D}_{1}}(3)$[GeV]  &\footnotesize    1797  &\footnotesize 628   &\footnotesize 1465 &\footnotesize  1445      &\footnotesize    393\\[-1.5mm]
\footnotesize $m_{\tilde{D}_{2}}(3)$[GeV]   &\footnotesize    1156  &\footnotesize 1439  &\footnotesize 2086 &\footnotesize 2059       &\footnotesize    1617\\[-1.5mm]
\footnotesize $\mu_D(3)$[GeV]                &\footnotesize    1466  &\footnotesize 1028  &\footnotesize 1747 &\footnotesize 1747      &\footnotesize    1055\\[-1.5mm]
\footnotesize $m_{\tilde{D}_{1}}(1,2)$[GeV]  &\footnotesize    1797  &\footnotesize 628   &\footnotesize 520   &\footnotesize 370             &\footnotesize    393 \\[-1.5mm]
\footnotesize $m_{\tilde{D}_{2}}(1,2)$[GeV]  &\footnotesize    1156  &\footnotesize 1439  &\footnotesize 906   &\footnotesize  916        &\footnotesize    1617 \\[-1.5mm]
\footnotesize $\mu_D(1,2)$[GeV]              &\footnotesize    1466  &\footnotesize 1028  &\footnotesize 300  &\footnotesize   391         &\footnotesize    1055 \\
\hline
\footnotesize $|m_{\chi^0_6}|$[GeV]          &\footnotesize    1278  &\footnotesize  1052 &\footnotesize 1054 &\footnotesize 1051       &\footnotesize    1203\\[-1.5mm]
\footnotesize $m_{h_3}\simeq M_{Z'}$[GeV]    &\footnotesize    1248  &\footnotesize  1020 &\footnotesize 1021 &\footnotesize 1021        &\footnotesize    1172\\[-1.5mm]
\footnotesize $|m_{\chi^0_5}|$[GeV]          &\footnotesize    1220  &\footnotesize  993  &\footnotesize 992 &\footnotesize 994     &\footnotesize    1143 \\
\hline
\footnotesize $m_S(1,2)$[GeV]                &\footnotesize    1097  &\footnotesize 908   &\footnotesize 1001 &\footnotesize 961       &\footnotesize    1093 \\[-1.5mm]
\footnotesize $m_{H_2}(1,2)$[GeV]            &\footnotesize    468  &\footnotesize 479   &\footnotesize 627  &\footnotesize 561     &\footnotesize    704 \\[-1.5mm]
\footnotesize $m_{H_1}(1,2)$[GeV]            &\footnotesize    165   &\footnotesize 154   &\footnotesize 459  &\footnotesize 345     &\footnotesize    220 \\[-2mm]
\footnotesize $\mu_{\tilde{H}}(1,2)$[GeV]    &\footnotesize    249  &\footnotesize 244   &\footnotesize 233  &\footnotesize 229        &\footnotesize    298 \\
\hline
\footnotesize $m_{\tilde{u}_1}(1,2)$[GeV]      &\footnotesize    893  &\footnotesize 788 &\footnotesize 911 &\footnotesize 845       &\footnotesize    929 \\[-1.5mm]
\footnotesize $m_{\tilde{d}_1}(1,2)$[GeV]      &\footnotesize    910  &\footnotesize 807 &\footnotesize 929  &\footnotesize 862      &\footnotesize    945 \\[-1.5mm]
\footnotesize $m_{\tilde{u}_2}(1,2)$[GeV]      &\footnotesize    910  &\footnotesize 807 &\footnotesize 929  &\footnotesize 862      &\footnotesize    945 \\[-1.5mm]
\footnotesize $m_{\tilde{d}_2}(1,2)$[GeV]      &\footnotesize    975  &\footnotesize 850 &\footnotesize 964 &\footnotesize 903         &\footnotesize    998 \\[-1.5mm]
\footnotesize $m_{\tilde{e}_2}(1,2,3)$[GeV]    &\footnotesize    874  &\footnotesize 733 &\footnotesize 849  &\footnotesize 796        &\footnotesize    900 \\[-1.5mm]
\footnotesize $m_{\tilde{e}_1}(1,2,3)$[GeV]    &\footnotesize     762  &\footnotesize 631 &\footnotesize 765  &\footnotesize 708        &\footnotesize    804 \\[-1.5mm]
\footnotesize $m_{\tilde{b}_2}$[GeV]         &\footnotesize    974  &\footnotesize 841   &\footnotesize 955  &\footnotesize 894     &\footnotesize    890 \\[-1.5mm]
\footnotesize $m_{\tilde{b}_1}$[GeV]         &\footnotesize     758  &\footnotesize 668   &\footnotesize 777  &\footnotesize 712      &\footnotesize    694 \\[-1.5mm]
\footnotesize $m_{\tilde{t}_2}$[GeV]         &\footnotesize     821  &\footnotesize 734   &\footnotesize 829  &\footnotesize 772      &\footnotesize    773 \\[-1.5mm]
\footnotesize $m_{\tilde{t}_1}$[GeV]         &\footnotesize     493  &\footnotesize 433   &\footnotesize 546   &\footnotesize 474     &\footnotesize    463 \\
\hline
\footnotesize $|m_{\chi^0_3}|\simeq |m_{\chi^0_4}|\simeq
|m_{\chi^{\pm}_2}|$[GeV]
                                             &\footnotesize     832  &\footnotesize 684   &\footnotesize 674      &\footnotesize  685        &\footnotesize    803 \\ [-1.5mm]
\footnotesize $m_{h_2}\simeq m_A \simeq m_{H^{\pm}}$[GeV]  &\footnotesize     615  &\footnotesize 664   &\footnotesize 963      &\footnotesize 720  &\footnotesize    593 \\[-1.5mm]
\footnotesize $m_{h_1}$[GeV]                 &\footnotesize     114  &\footnotesize 115   &\footnotesize 115     &\footnotesize 114   &\footnotesize    119 \\ \hline
\footnotesize $m_{\tilde{g}}$[GeV]           &\footnotesize     336  &\footnotesize 330   &\footnotesize 353     &\footnotesize 327    &\footnotesize    338 \\[-1.5mm]
\footnotesize $|m_{\chi^{\pm}_1}|\simeq |m_{\chi^0_2}|$[GeV]
                                             &\footnotesize     107  &\footnotesize 103   &\footnotesize 109      &\footnotesize 101   &\footnotesize    103 \\[-1.5mm]
\footnotesize $|m_{\chi^0_1}|$[GeV]          &\footnotesize  59 &\footnotesize 58    &\footnotesize 61     &\footnotesize 57  &\footnotesize    58  \\
\hline
\end{tabular}
\caption{The ``early discovery'' cE$_6$SSM benchmark points.}
\label{table:benchmarks}
\end{center}
\end{table}

\end{document}